\documentstyle[10pt,psfig]{espart}
\begin{document}
\begin{frontmatter}
\title{{\small In Hwa~A. Lim und Charles~A. Cantor, editors, {\em Proc. 3. Int.
  Conf. on\\[-0.3cm] Bioinformatics and Genome Research}, page 445,
(World Scientific, 1994).}\\[1.5cm]
  On the Hypercube Structure of \\the Genetic Code}
\author[Puebla]{Miguel A. Jim\'enez--Monta\~no\thanksref{bylinejimm}}
\author[Xalapa]{Carlos R. de la Mora--Bas\'a\~nez\thanksref{bylinecarlos}}
\author[Berlin,Paris]{Thorsten P\"oschel\thanksref{bylinetp}}

\address[Puebla]{Departamento de F\'{\i}sica y Matem\'aticas, Universidad de
las Am\'ericas, Puebla, Sta. Catarina M\'artir, 72820 Cholula, Puebla,
M\'exico}
\address[Xalapa]{Direcci\'on General de Investigaciones, Universidad 
Veracruzana, Xalapa, Ver. 91000, M\'exico}
\address[Berlin]{Humboldt--Universit\"at zu Berlin, Institut f\"ur
Theoretische Physik, Invalidenstra\ss e 110, D--10099 Berlin, Germany} 
\address[Paris] {Laboratoire de Physique et Mechanique des Milieux
Heterogenes, Ecole Superieure de Physique et de Chimie Industrielles,
10 rue Vauquelin, F--75231 Paris Cedex 05, France}

\thanks[bylinejimm]{E--mail: jimm@udlapvms.pue.udlap.mx}
\thanks[bylinecarlos]{E--mail: delamora@speedy.coacade.uv.mx}
\thanks[bylinetp]{E--mail: thorsten.poeschel@charite.de}

\begin{abstract}
 A representation of the genetic code as a six--dimensional Boolean
hypercube is proposed. It is assumed here that this structure is the
result of the hierarchical order of the interaction energies of the
bases in codon--anticodon recognition. The proposed structure
demonstrates that in the genetic code there is a balance between
conservatism and innovation. Comparing aligned positions in homologous
protein sequences two different behaviors are found: a)There are sites
in which the different amino acids present may be explained by one or
two ``attractor nodes'' (coding for the dominating amino acid(s)) and
their one--bit neighbors in the codon hypercube, and b) There are
sites in which the amino acids present correspond to codons located in
closed paths in the hypercube. The structure of the code facilitates
evolution: the variation found at the variable positions of proteins
do not corresponds to random jumps at the codon level, but to well
defined regions of the hypercube.
\end{abstract}
\end{frontmatter}

\section{Introduction}
The genetic code is the biochemical system for gene expression. It
deals with the translation, or decoding, of information contained in
the primary structure of DNA and RNA molecules into protein sequences.
Therefore, the genetic code is both a physico--chemical and a
communication system. Physically molecular recognition depends on the
degree of complementarity between the interacting molecular surfaces
(by means of weak interactions); informationally, a prerequisite to
define a code is the concept of distinguishability. It is the physical
indistinguishability of some codon--anticodon interaction energies that
makes the codons synonymous, and the code degenerate and redundant
\cite{1}.

In natural languages \cite{2} as well as in the genetic code, the total
redundancy is due to a hierarchy of constraints acting one upon
another. The specific way in which the code departs from randomness
is, by definition, its structure. It is assumed here that this
structure is the result of the hierarchical order of the interaction
energies of the bases in codon--anticodon recognition. As we shall see,
it may be represented by a six--dimensional boolean hypercube in which
the codons (actually the code--words; see below) occupy the vertices
(nodes) in such a way that all kinship neighborhoods are correctly
represented. This approach is a particular application to binary
sequences of length six of the general concept of sequence--space,
first introduced in coding theory by Hamming \cite{3}. 

A code--word is next to six nodes representing codons differing in a
single property. Thus the hypercube simultaneously represents the
whole set of codons and keeps track of which codons are one--bit
neighbors of each other. Different hyperplanes correspond to the four
stages of the evolution of the code according to the Co--evolution
Theory \cite{4,5,6}. Hops within three of the ``columns'' (four--dimensional
cubes), consisting of the codon classes NGN, NAN, NCN, and NUN, lead
to silent and conservative amino acid substitutions, while hops in
the same hyperplane (four--dimensional subspace belonging to any of the
codon classes ANN, CNN, GNN or UNN) lead to non--conservative
substitutions, frequently found in proteins. The proposed structure
demonstrates that in the genetic code there is a good balance between
conservatism and innovation. To illustrate the results several
examples of the non--conservative variable positions of homologous
proteins are discussed. Two different behaviors are found: a) There
are sites in which the different amino acids present may be explained
by one or two ``attractor nodes'' (coding for the dominating amino
acid(s)) and their one--bit neighbors in the codon hypercube, and b)
There are sites in which the amino acids present correspond to codons
located in closed paths in the hypercube. 

\section{Codon--Anticodon Interaction}
In his early paper Eigen \cite{7} recognized that the optimization
between stability and rate, that is always found for enzyme--substrate
interactions, also applies to the codon--anticodon interaction.
However, he attributes the codon압 size to mechanistic coincidences:
``codons with less than three bases would be very unstable (at least
for A and U). Codons with more than three bases, especially for G and
C, become too `sticky' ''. This is certainly not a coincidence, but a
requirement for the system to function as an efficient communication
device. Three bases are needed for effectively binding the adapter to
the messanger. Thus, the codon압 size determines the range of
codon--anticodon overall interaction strength within which recognition
can occur\footnote{Interestingly enough, this feature of genetic
communication system has, its counterpart in human communication. In a
series of experiments on reading lists of words, performed by J.E.
Karlin and J.R. Pierce (Pierce J.R., An Introduction to Information
Theory, Dover Publications,Inv. N.Y. 1961), in which the subject
``transmits'' the information translating it into the new form, speech
rather than print, by reading the list aloud, they concluded that:
``It seems fairly clear that reading speed is limited by
\underline{word recognition} not by \underline{word utterance}''
(underlined in the original). See also \cite{12}}. Genetic translation
rate is limited, among other things, by codon--anticodon recognition
which in turn depends on base--pair lifetimes in a given structural
situation. These life--times are influenced by the nature of the pairs:
they are shorter for AT than for GC pairs \cite{8}.

The four bases occurring in DNA (RNA) macromolecules define the
corresponding alphabet X: \{A, C, G, T\} or \{A, C, G, U\}. Each base
is completely specified by two independent dichotomic
categorizations (Fig. \ref{fig1}):
\begin{figure}
\centerline{\psfig{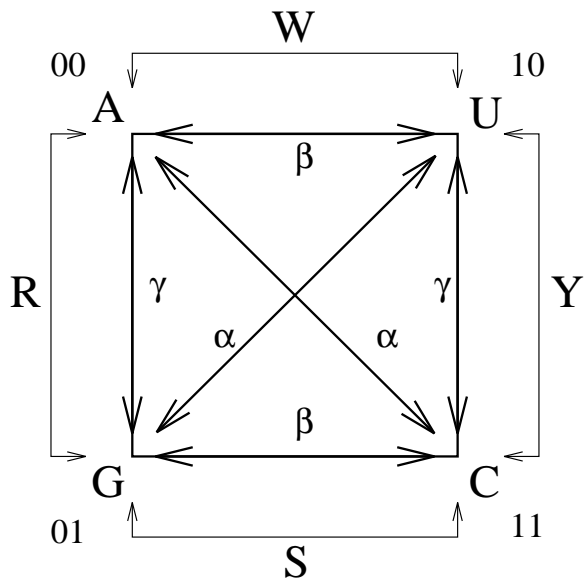}}
\caption{Categorizations of the bases. The categorizations of the
bases according to (i): chemical type C : \{R, Y\}, where R: (A, G)
are purines and Y: (C, U) are pyrimidines, and (ii) according to
H--bonding, H : \{W, S\}, where W: (A, U) are weak and S: (C, G)
strong bases. The third possible partition into imino/keto bases is
not independent on the former ones and is irrelevant for the
codon--anticodon interaction. The binary representation of the bases
is also shown. The first bit is the chemical type and the second one
the H--bonding character. $\alpha$, $\beta$ and $\gamma$ are the
transformations of the bases, which form a Klein--4 group [6,8].}
\label{fig1}
\end{figure}

(i) according to chemical type C : \{R, Y\}, where R: (A, G) are
purines and Y:(C, U) are pyrimidines, and (ii) according to
H--bonding, H : \{W, S\}, where W:(A, U) are weak and S:(C, G) strong
bases. The third possible partition into imino/keto bases is not
independent from the former ones. 

Denoting by $C_i$ the chemical type and by $H_i$ the $H$--bond
category of the base $B_i$, at position $i$ of a codon, our basic
assumption says that the codon--anticodon interaction energy obeys the
following hierarchical order:
\begin{eqnarray}
&&C_2 > H_2 > C_1 > H_1 > C_3 > H_3\nonumber
\end{eqnarray}
This means, that the most important characteristic determining the
codon--anticodon interaction is the chemical type of the base in the
second position. The next most important characteristic is whether
there is a weak or strong base in this position, then the chemical
type of the first base and so on.

The bases are represented by the nodes of a 2--cube (Fig. \ref{fig1}).
The first attribute is the chemical character and the second the
hydrogen--bond character. Extending this association to base triplets,
each codon is in a unique way associated with a codeword consisting of
six attribute values (see Table \ref{tab1}).
\begin{table}[ht]
\renewcommand{\arraystretch}{0.8}
\footnotesize
\centerline{\begin{tabular}{|c c c c c c | c c c | c |} \hline
0 & 0 & 0 & 0 & 1 & 1 & A & A & C & N \\
0 & 0 & 0 & 0 & 1 & 0 & A & A & U & N \\
0 & 0 & 0 & 0 & 0 & 0 & A & A & A & K \\
0 & 0 & 0 & 0 & 0 & 1 & A & A & G & K \\
1 & 0 & 0 & 0 & 0 & 1 & U & A & G & t \\
1 & 0 & 0 & 0 & 0 & 0 & U & A & A & t \\
1 & 0 & 0 & 0 & 1 & 0 & U & A & U & Y \\
1 & 0 & 0 & 0 & 1 & 1 & U & A & C & Y \\
1 & 1 & 0 & 0 & 1 & 1 & C & A & C & H \\
1 & 1 & 0 & 0 & 1 & 0 & C & A & U & H \\
1 & 1 & 0 & 0 & 0 & 0 & C & A & A & Q \\ 
1 & 1 & 0 & 0 & 0 & 1 & C & A & G & Q \\
0 & 1 & 0 & 0 & 0 & 1 & G & A & G & E \\
0 & 1 & 0 & 0 & 0 & 0 & G & A & A & E \\
0 & 1 & 0 & 0 & 1 & 0 & G & A & U & D \\
0 & 1 & 0 & 0 & 1 & 1 & G & A & C & D \\
0 & 1 & 1 & 0 & 1 & 1 & G & U & C & V \\
0 & 1 & 1 & 0 & 1 & 0 & G & U & U & V \\
0 & 1 & 1 & 0 & 0 & 0 & G & U & A & V \\
0 & 1 & 1 & 0 & 0 & 1 & G & U & G & V \\
1 & 1 & 1 & 0 & 0 & 1 & C & U & G & L \\
1 & 1 & 1 & 0 & 0 & 0 & C & U & A & L \\
1 & 1 & 1 & 0 & 1 & 0 & C & U & U & L \\
1 & 1 & 1 & 0 & 1 & 1 & C & U & C & L \\
1 & 0 & 1 & 0 & 1 & 1 & U & U & C & F \\
1 & 0 & 1 & 0 & 1 & 0 & U & U & U & F \\
1 & 0 & 1 & 0 & 0 & 0 & U & U & A & L \\
1 & 0 & 1 & 0 & 0 & 1 & U & U & C & L \\
0 & 0 & 1 & 0 & 0 & 1 & A & U & G & M \\
0 & 0 & 1 & 0 & 0 & 0 & A & U & A & I \\
0 & 0 & 1 & 0 & 1 & 0 & A & U & U & I \\
0 & 0 & 1 & 0 & 1 & 1 & A & U & C & I \\ \hline
\end{tabular}\hspace{0.5cm}
\begin{tabular}{|c c c c c c | c c c | c |} \hline
0 & 0 & 1 & 1 & 1 & 1 & A & C & C & T \\
0 & 0 & 1 & 1 & 1 & 0 & A & C & U & T \\
0 & 0 & 1 & 1 & 0 & 0 & A & C & A & T \\
0 & 0 & 1 & 1 & 0 & 1 & A & C & G & T \\
1 & 0 & 1 & 1 & 0 & 1 & U & C & G & S \\
1 & 0 & 1 & 1 & 0 & 0 & U & C & A & S \\
1 & 0 & 1 & 1 & 1 & 0 & U & C & U & S \\
1 & 0 & 1 & 1 & 1 & 1 & U & C & C & S \\
1 & 1 & 1 & 1 & 1 & 1 & C & C & C & P \\
1 & 1 & 1 & 1 & 1 & 0 & C & C & U & P \\
1 & 1 & 1 & 1 & 0 & 0 & C & C & A & P \\
1 & 1 & 1 & 1 & 0 & 1 & C & C & G & P \\
0 & 1 & 1 & 1 & 0 & 1 & G & C & G & A \\
0 & 1 & 1 & 1 & 0 & 0 & G & C & A & A \\
0 & 1 & 1 & 1 & 1 & 0 & G & C & U & A \\
0 & 1 & 1 & 1 & 1 & 1 & G & C & C & A \\
0 & 1 & 0 & 1 & 1 & 1 & G & G & C & G \\
0 & 1 & 0 & 1 & 1 & 0 & G & G & U & G \\
0 & 1 & 0 & 1 & 0 & 0 & G & G & A & G \\
0 & 1 & 0 & 1 & 0 & 1 & G & G & G & G \\
1 & 1 & 0 & 1 & 0 & 1 & C & G & G & R \\
1 & 1 & 0 & 1 & 0 & 0 & C & G & A & R \\
1 & 1 & 0 & 1 & 1 & 0 & C & G & U & R \\
1 & 1 & 0 & 1 & 1 & 1 & C & G & C & R \\
1 & 0 & 0 & 1 & 1 & 1 & U & G & C & C \\
1 & 0 & 0 & 1 & 1 & 0 & U & G & U & C \\
1 & 0 & 0 & 1 & 0 & 0 & U & G & A & t \\
1 & 0 & 0 & 1 & 0 & 1 & U & G & C & W \\
0 & 0 & 0 & 1 & 0 & 1 & A & G & G & R \\
0 & 0 & 0 & 1 & 0 & 0 & A & G & A & R \\
0 & 0 & 0 & 1 & 1 & 0 & A & G & U & S \\
0 & 0 & 0 & 1 & 1 & 1 & A & G & C & S \\ \hline
\end{tabular}}
\caption{Gray code representation of the genetic code. In the first
and fourth blocks the 6--binary vectors (code--words) are shown. In the
second and fifth blocks appear the corresponding codons. Finally in
the third and sixth columns the amino acids in single letter notation.
The first two digits correspond to the first base, the following two
to the second base and the last two to the last base, according to the
binary codification of the bases of Fig. \ref{fig1}}
\label{tab1}
\end{table}

\begin{figure}[ht]
\centerline{\psfig{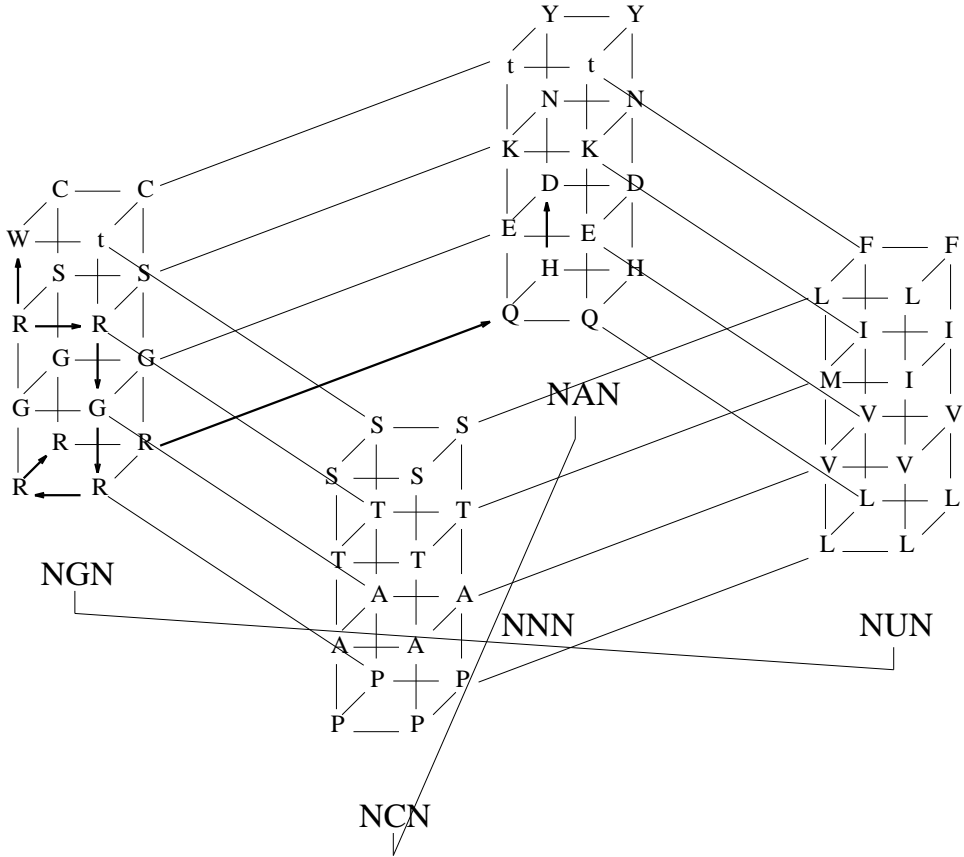}}
\caption{The six--dimensional hypercube. Each node is labeled with
the corresponding amino acid, in the single letter notation, or
terminator symbol. For clarity, only some of the links are shown. The
cluster of amino acids of the first example discussed in the text and
their links are displayed.}
\label{fig2a}
\end{figure}

In some of the hypercube directions single feature codon changes
(one--bit code--word changes) produce synonymous or conservative amino
acid substitutions in the corresponding protein (when the hops occur
in three of the 4--cubes displayed as ``columns'' in Figs. \ref{fig2a} and
\ref{fig2b}) while in other directions lead to context dependent
replacements which in general conserve only certain physical
properties. However, if these properties are the only relevant ones in
the given context, the substitution has little effect on the protein
structure as well. These low--constraint sites facilitate evolution
because they allow the transit between hypercube columns belonging to
amino acids with very different physico--chemical properties
(e.g. hydrophobic and hydrophilic amino acids, respectively).

\section{Gray Code Structure of the Genetic Code}
\begin{figure}[ht]
\centerline{\psfig{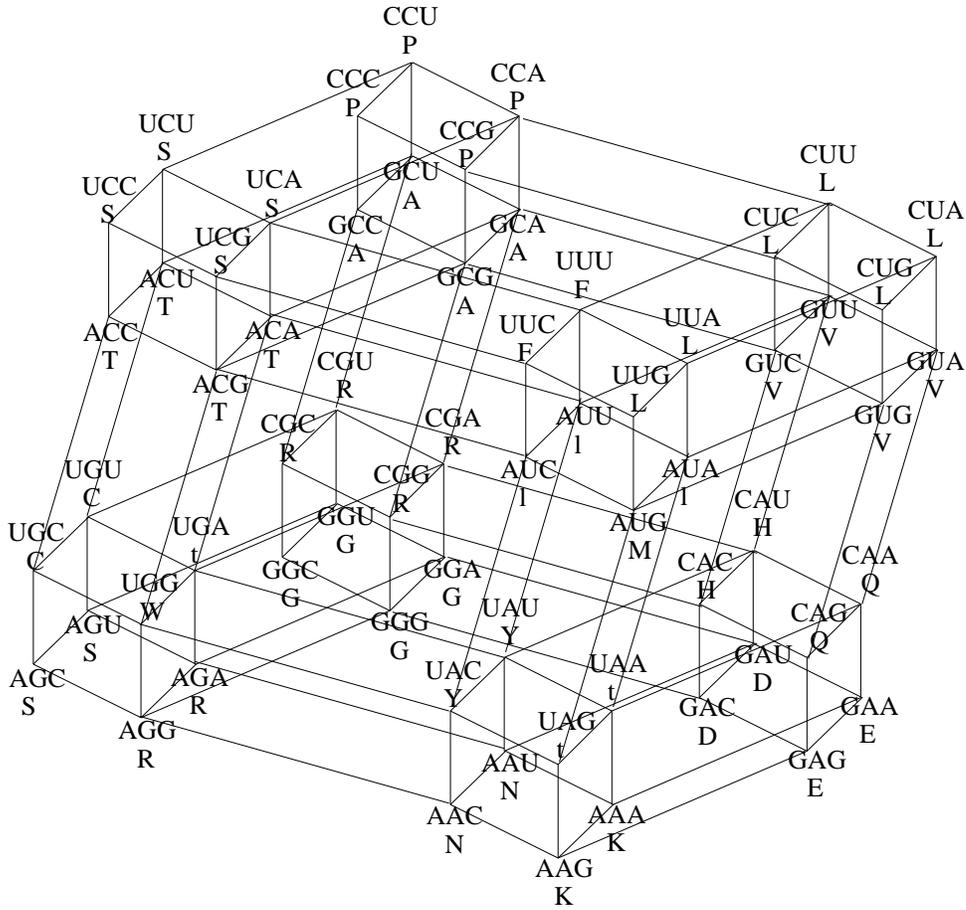}}
\caption{The hypercube representation of the genetic code. Each node
represents a code--word (6--binary vector) of attribute
values. However, for clarity of interpretation, the nodes are labeled
with the corresponding codons (See Table \ref{tab1} for the assignment of
codons to vectors). The nodes and links mentioned in second example
discussed in the text are shown.}
\label{fig2b}
\end{figure}

An $n$--dimensional hypercube, denoted by $Q_n$, consists of $2^n$
nodes, each addressed by a unique $n$--bit identification number. A
link exists between two nodes of $Q_n$ if and only if their node
addresses differ in exactly one bit position. A link is said to be
along dimension $i$ if it connects two nodes whose addresses differ to
as the $i$th bit (where the least significant bit is referred to as
the 0th bit). $Q_6$ is illustrated in Fig. \ref{fig2b}. Two nodes in
a hypercube are said to be adjacent if there is a link present between
them. The (Hamming) distance between any two cube nodes is the number
of bits differing in their addresses. The number of hops needed to
reach a node from another node equals the distance between the two
nodes. A $d$--dimensional subcube in $Q_n$ involves $2^d$ nodes whose
addresses belong to a sequence of n symbols \{0, 1, *\} in which exactly
$d$ of them are of the symbol ``*'' (i.e. they don앖 care symbol whose
value can be 0 or 1).

The idea to propose a Gray Code representation of the Genetic Code
goes back to Swanson \cite{9} where this concept is explained in
detail (see also \cite{10}). A great number of different Gray Codes
can be associated to the Genetic Code, depending of the order of
importance of the bits in a code--word. In Table \ref{tab1} our chosen
Gray Code is displayed. It is constructed according to our main
hypothesis 
\begin{eqnarray}
C_2 > H_2 > C_1 > H_1 > C_3 > H_3 \nonumber
\end{eqnarray}
For example, the first two lines of the table differ in the last bit,
corresponding to $H_3$ which is the least significant bit; the second
and the third lines differ in the next least significant bit, i.e.
$C_3$, and so forth. 

\section{The Structure of Codon Doublets}
This section is more mathematical than the rest of the paper,
therefore it is suggested to non--mathematical readers to skip the
details. This will not be an obstacle for the understanding of the
rest of the paper.

\begin{figure}[ht]
\centerline{\psfig{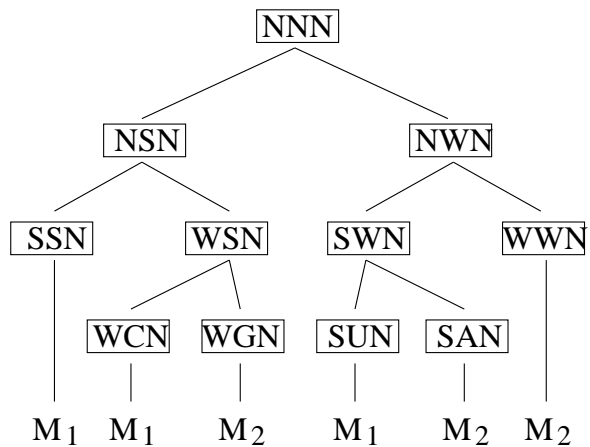}}
\caption{Decision--tree of codon categories and redundancy
distribution. The leaves are the sets of four--fold ($M_1$) and less
than four--fold ($M_2$) degenerate $B_1B_2$ doublets.}
\label{fig3}
\end{figure}

In a pioneering paper Danckwerts and Neubert \cite{11} discussed the
symmetries of the sixteen $B_1B_2$ codon doublets in terms of the Klein--4
group of base transformations. Here their result will be recast in a
form of a decision--tree (Fig. \ref{fig3}), and their analysis will be
extended to the $B_2B_3$ doublets.

They found the following structure for the set $M$ of $B_1B_2$
doublets: 

Starting from AC generate the set:
\begin{eqnarray}
M_0 &=& \{[(1,1) \cup (\alpha, 1) \cup (\alpha, \beta) \cup (\alpha,
\gamma)] AC \} = \{ AC, CC, CG, CU \} \nonumber \\
M_1 &=& [(1,1) \cup (\beta, 1)] M_0 \hspace{0.5cm}\mbox{ and}\nonumber \\ 
M_2 &=& (\alpha, \alpha) M_1 \nonumber
\end{eqnarray}
The sets $M_1$ and $M_2$ consist of four--fold and less than four--fold
degenerate doublets, respectively. 

The set M can be expressed as:
\begin{eqnarray}
M = [(1,1) \cup (\beta, 1)]~[(1,1) \cup (\alpha, \alpha)] M_0 \nonumber 
\end{eqnarray}
Where the base exchange operators ($\alpha$, $\beta$, $\gamma$) are
defined in Fig. \ref{fig1}. 

They showed that: ``$M_1$ and $M_2$ are invariant by operating with
$(\beta, 1)$ on $B_1$, but no operation on $B_2$ leaves $M_1$ or $M_2$
invariant. Thus $B_2$ carries more information\footnote{see also
\cite{12}} than $B_1$ and $B_2$ is therefore more important for the
stability of $M_1$ and $M_2$ than $B_1$. A change of $B_1$ with
respect to its hydrogen bond property does not change the resulting
amino acids if all doublets of either $M_1$ or $M_2$ are affected.

Reversing supposition and conclusion, $M_1$ and $M_2$ may be defined
as those doublet sets of 8 elements which are invariant under the
$(\beta, 1)$--transformation. Then experience shows that $M_1$ and
$M_2$ are fourfold and less than fourfold degenerate respectively.''

Thus, the third base degeneracy of a codon does not depend on the
exact base $B_1$, but only on its $H$--bond property (weak or strong).

 The above results can be simply visualized as a decision--tree (Fig.
\ref{fig3}). As can be seen from this figure, the redundancy of a
codon is determined only by the H--bond character of $B_1$ and $B_2$:
SSN codons (with 6 H--bonds in $B_1B_2$) belong to $M_1$ while WWN
codons (with 4 H--bonds in $B_1B_2$) belong to $M_2$. However, for
codons WSN and SWN (with 5 H--bonds in $B_1B_2$) it is not possible to
decide unless one has more information about the second base: WCN and
SUN belong to $M_1$ while WGN and SAN belong to $M_2$. In all cases
at most three attributes are necessary to determine the redundancy of
a codon up to this point, of course, the non--degenerate codons (UAG
for Methionine and UGG for Thryptophan) will require the specification
of the six attributes.

From the decision rules obtained from Fig. \ref{fig3} it is clear that
there are branches where the refinement procedure cannot continue (the
branches which end in $M_1$) because no matter which base occupies
the third codon position the degeneracy cannot be lifted. This imposes
a limit to the maximum number of amino acids which can be incorporated
to the code without recurring to a ``frozen accident'' hypothesis. Our
proposal generalizes the ``2--out--of--3'' hypothesis of Lagerkvist
\cite{12}, which refers only to codons in the SSN class.

\begin{figure}[ht]
\centerline{\psfig{figure=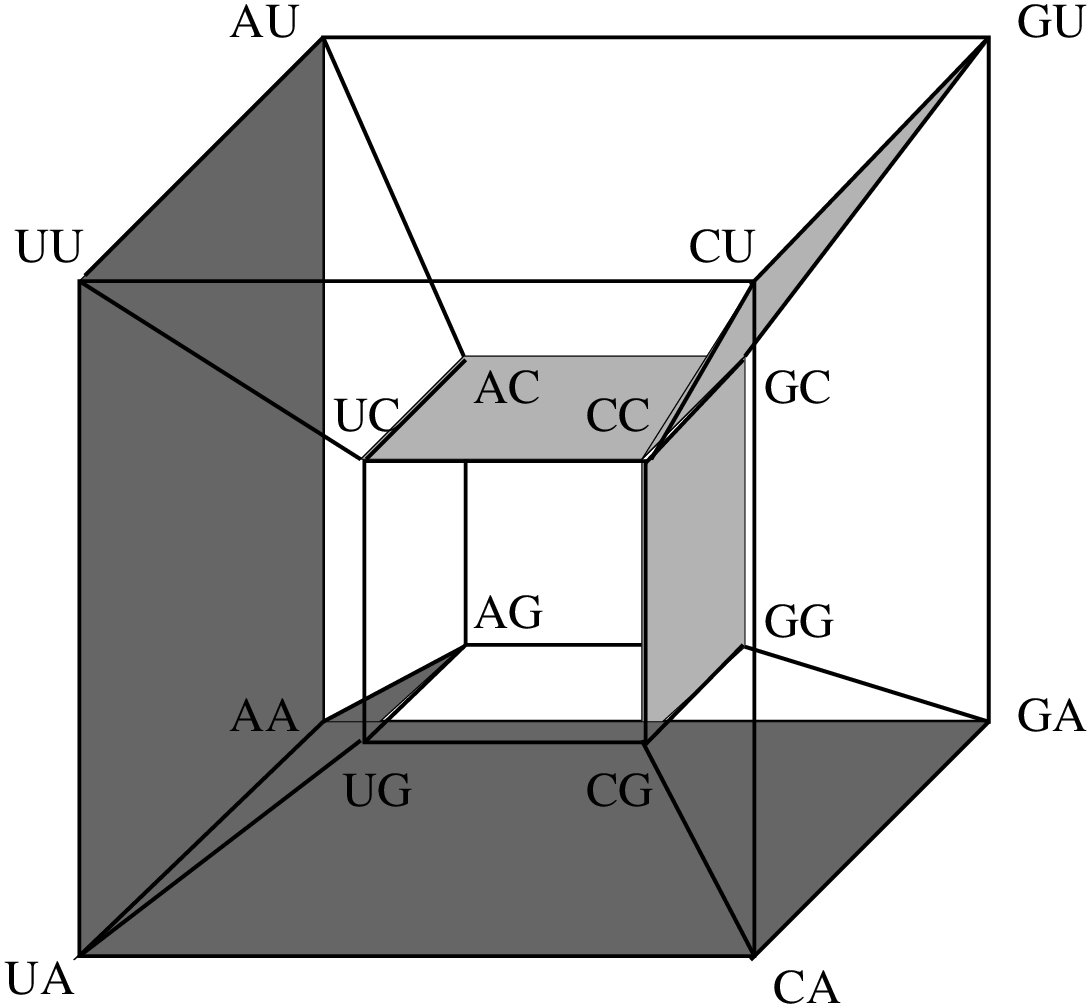,width=3.0in}}
\caption{The four--dimensional hypercube representation of the sets
$M_1$ and $M_2$.}
\label{fig4a}
\end{figure}

The sixteen $B_1B_2$ doublets can be represented as the vertices of a
four--dimensional hypercube (Fig. \ref{fig4a}). As can be seen from this
figure, the sets $M_1$ and $M_2$ are located in compact regions.
Notice that this figure differs from the one introduced by Bertman and
Jungck \cite{13}, which considered as basic transformations $\alpha$
and $\beta$ instead of $\beta$ and $\gamma$ as we did. Since the
operator $\alpha$ changes two bits we do not consider it as basic.

Let압 consider now the structure of the set $M'$ of $B_2B_3$ doublets.
Exactly as before, define the following sets: 
\begin{eqnarray}
M_0' &=& \{NC\} \nonumber \\
M_1' &=& [(1,1) \cup (1,\beta)] M_0' \hspace{0.5cm}\mbox{ and}\nonumber \\
M_2' &=& (\alpha, \alpha) M_1' \hspace{0.5cm}\mbox{ (alternatively } M_1'=(\alpha,
\alpha) M_2' \mbox{ )}\nonumber
\end{eqnarray}
where $M_1'$ consists of the doublets $B_2B_3$ ending in a strong base
(NS), and $M_2'$ of the doublets ending in a weak base (NW). Then
\begin{eqnarray}
M' &=& M_1' \cup M_2' \hspace{0.5cm}\mbox{ can be expressed as} \nonumber \\
M' &=& [(1,1) \cup (1,\beta)]~[(1,1) \cup (\alpha, \alpha)] M_0'
\nonumber
\end{eqnarray}
Notice that the operator acting on $M_0'$ has the same form as the
operator acting on $M_0$ above, except that b acts as the third base
instead of the first. 

The sets $M_1'$ and $M_2'$ are invariant under the (1,
b)--transformations. Then experience shows that the 32 codons in the
class $NB_2B_3$, with $B_2B_3$ in $M_1'$ or $M_2'$ constitute a
complete code, codifying for the 20 amino acids and terminator signal
(stop--codon) if allowance is made for deviating codon--assignments
found in Mitochondria \cite{14}. For the codons in $M_1'$ this is true
in the universal code; for codons in $M_2'$ AUA should codify for M
instead of I and UGA for W instead of stop signal. Both changes have
been observed in Mitochodria. This more symmetric code has been
considered more similar to an archetypal code than the universal code
\cite{14}. Only after the last attribute $H_3$ was introduced the
universal code was obtained with the split of AUR into AUA (I) and AUG
(M); and UGR into UGG (W) and UGA (t). It has been speculated that
primordial genes could be included in a 0.55--kb open reading frame
\cite{15}. The same authors calculated that with two stop codons this
open reading frames would have appeared too frequently. From the
present view the assignment of UGA to a stop codon was a late event
that optimized this frequency (this interpretation differs from the
one proposed in
\cite{15} and \cite{16} which assume a primordial code with three stop
codons). Other deviations of the universal code most likely also
occurred in the last stages of the code압 evolution.

\begin{figure}[ht]
\centerline{\psfig{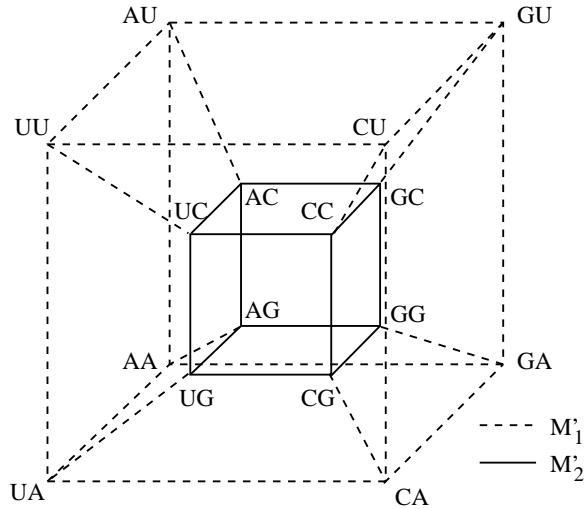}}
\caption{The corresponding to fig. \ref{fig4a} hypercube of the sets
$M_1'$ and $M_2'$. Notice that in both cases each set is located in
a compact region.}
\label{fig4b}
\end{figure}

In the same way as before the sixteen $B_2B_3$ doublets can be
represented as the vertices of a four--dimensional hypercube
(fig. \ref{fig4b}). The sets $M_1'$and $M_2'$ are also located in compact
regions. Codons with $B_2B_3$ in $M_1'$ are frequently used in eukaryots.
In contrary, codons with $B_2B_3$ in $M_2'$ are frequently used in
prokaryots. The described structure of the code allows a modulation of
the codon--anticodon interaction energy \cite{17}.

\section{Results}
\begin{figure}[ht]
\centerline{\psfig{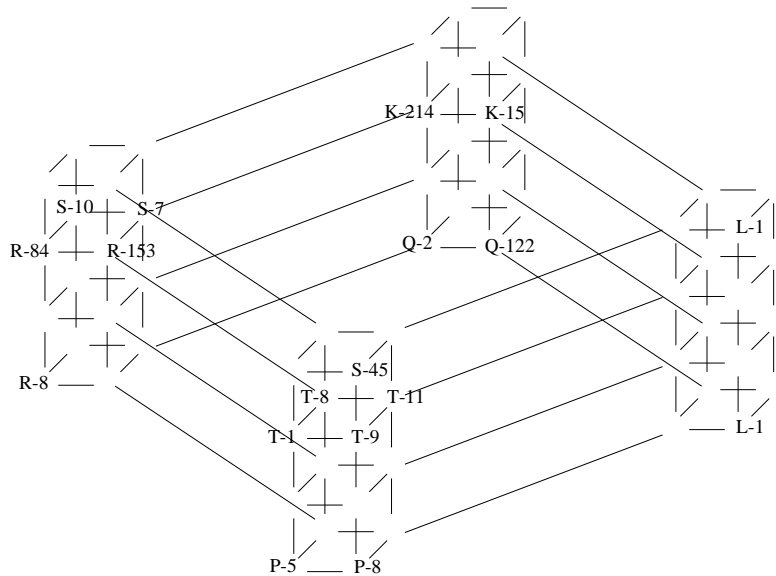}}
\caption{The amino acid hypercube with the amino acids at position 18
of the variable region of kappa light chain displayed. The number
after the amino acid symbol is the number of times the amino acid
occurs in the alignment of Kabat et al. (1991): Sequences of proteins
of immunological interest, 5th ed. NIH, Bethesda, MD.} 
\label{fig5}
\end{figure}

Besides the results mentioned in the last section which refer to codon
doublets, to further illustrate the significance of proposed approach
we are going to consider several examples. In the first example
(Fig. \ref{fig5}) we discuss the alignment studied by the method of
hierarchical analysis of residue conservation of Livingstone and
Barton (Fig. 2 of \cite{18}). In position 11 appear the following
amino acids: R, W, H, G, D, which according to their approach have no
properties in common. In Fig \ref{fig2a} this cluster of amino acids
is shown.  By looking at the Atlas of amino acid properties \cite{19}
we see that, from the properties proposed by Grantham \cite{20}
(composition, polarity and volume), apparently the only requirement
for the amino acids at this site is to maintain a certain degree of
polarity. From this observation we may conclude that most probably it
is an external site. Simply by looking at such a diverse set of amino
acids one can hardly realize that they have clustered codons. This
clustering facilitates the occurrence of mutations that in the course
of evolution were fixed, in view of the low physico--chemical
requirements at the site.

As a second example (Fig. \ref{fig2b}) we consider site 33 of the
alignment of 67 SH2 domains, Fig 6 of \cite{8}. We can see from
Fig. \ref{fig2b} that the cluster around the codon CAC (H) explains, by
one--bit changes, the amino acids R, Q, L, H, D. Furthermore, a second
cluster around the codon AGC (S) explains the amino acids R, N, S,
T. Finally, a silent change from AGC (S) to UCC (S) accounts for the
minor appearance of the small neutral amino acids A, T, P. In a
similar way, the variation of the hypervariable region of
immunoglobulin kappa light FR1 at position 18 can be explained
(Fig. \ref{fig5}. Finally, by looking to the residue frequencies in
226 globins displayed in Table 3 of the paper by Bashford et
al. \cite{21} it is seen that there are variable positions in which
one or two residues predominately occur and the rest are only
marginally represented, and others in which the frequencies are more
evenly distributed among the amino acids present. As it can be easily
shown, the first class of positions may be associated at the codon
level with one (or two) attractor node(s) and its one--bit neighbors,
and the second one with closed trajectories in the hypercube. The
corresponding figures are not included because of lack of space.

\section{Concluding Remarks}
 The present approach goes beyond the usual analyses in terms of
single base changes because it takes into account the two characters
of each base and therefore it represents one--bit changes. Besides,
the base position within the codon is also considered. The fact that
single bit mutations occur frequently is expected from probabilistic
arguments. However, one could not expect, a priori, that a cluster of
mutations would correspond at the amino acid level to a cluster of
amino acids fixed by natural selection. We have found that this
situation presents itself for many positions of homologous protein
sequences of many different families (results not included). The
structure of the code facilitates evolution: the variations found at
the variable positions of proteins do not corresponds to random jumps
at the codon level, but to well defined regions of the hypercube.

\ack 
This work received economical support from: Proyecto CONACyT
No. 1932--E9211. We thank professors Werner Ebeling and Michael Conrad
for encouraging comments.

\end{document}